\documentclass[a4paper,11pt]{article}
\usepackage{pos}
\usepackage{braket}
\newcommand*\dif{\mathop{}\!\mathrm{d}}
\newcommand{\ar}{\arrowvert}
\newcommand{\ra}{\rangle}

\title{Vacuum replicas in field-theory models of Coulomb QCD}
\author[a]{Pedro J. de A. Bicudo}
\author*[b]{Eduardo Garnacho Velasco}
\author[b]{Felipe J. Llanes-Estrada}
\author[a]{\\ J. Emilio Ribeiro}
\author[b]{V\'{\i}ctor Serrano Herreros}
\author[b]{Jorge Vallejo Fern\'andez}

\affiliation[a]{Dep. Física and CeFEMA, Instituto Superior Técnico, Av. Rovisco Pais, 1049-001 Lisboa, Portugal}

\affiliation[b]{Univ. Complutense de Madrid, Dept. F\'{\i}sica Te\'orica, Plaza de las Ciencias 1, 28040 Madrid, Spain}

\emailAdd{fllanes@fis.ucm.es}
\abstract{Dynamical symmetry breaking can happen through a Higgs mechanism but also spontaneously within a strongly-enough coupled theory. We treat a field-theoretical quark model of QCD based on a linear+Coulomb Cornell potential (to account for the longitudinal interaction), together with a transverse interaction (to account for Coulomb-gauge gluons) in BCS approximation.
After extracting the well-known BCS ground state on which abundant hadron phenomenology
implementing dynamical chiral symmetry breaking has been built, we find two excited replica
vacuumlike states shown in figure 1. 
At the BCS level they are classically unstable, but second quantization blocks transitions between them in volumes much larger than a hadron size. Mesons built over the replica vacua have relative masses similar to normal mesons on the ordinary vacuum; we find no negative-mass mode, confirming their stability found earlier with a harmonic oscillator potential.
\centerline{\includegraphics[width=0.8\columnwidth]{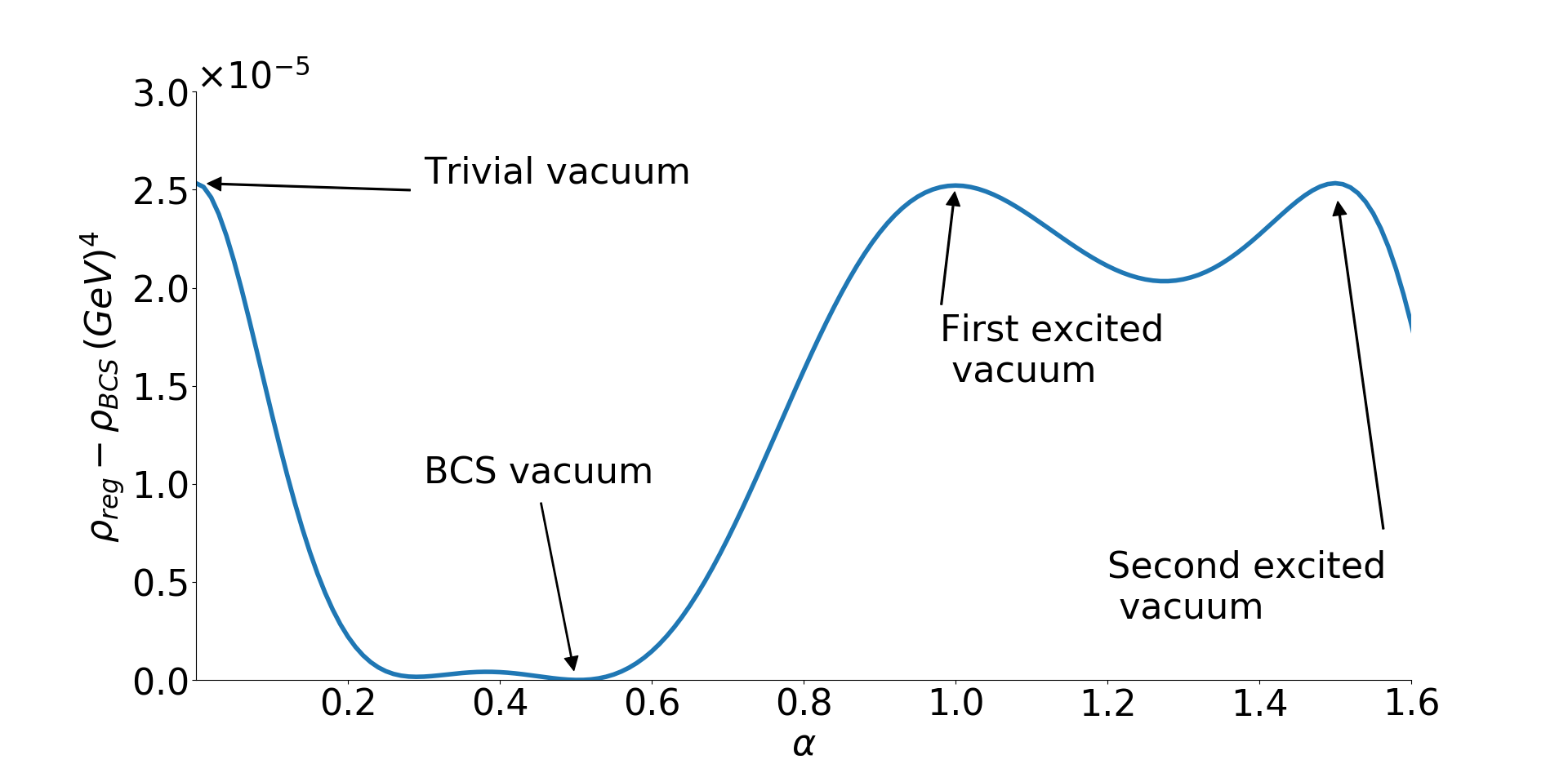}}
{\bf Figure 1:} $\langle H\rangle$ plot through Cooper-pair function space along a curve parametrized by $\alpha$\\
that passes by all relevant states described in the text.}

\FullConference{%
  *** The European Physical Society Conference on High Energy Physics (EPS-HEP2021), ***\\
  *** 26-30 July 2021 ***\\
  *** Online conference, jointly organized by Universität Hamburg and the research center DESY ***
}

\begin{document}
\maketitle
\setcounter{figure}{1} 

The Hamiltonian expectation value $\langle H\rangle$ of
field theories can have multiple local minima  (``vacua''). This happens independently of any topological degeneracy of the actual ground state (such as the $\theta$-vacua of QCD); the alternative minima that we here discuss can have more energy than the ground state but be trapped behind large enough (or infinite) barriers that hinder their decay. They can be thought as excited vacua or ``replicas'', a name appropriate because
entire additional Fock-spaces of excited physical particles can be built over them.

It has now been known for a time that replicas appear in many relativistic systems such as color-confining models of QCD~\cite{Bicudo:2002eu}, other field theories such as fermion+scalar ones~\cite{Llanes-Estrada:2006bxu} or extended space-times~\cite{Colin-Ellerin:2020mva}, etc.
Recent work~\cite{Bicudo:2019ryc} has focused in showing that the QCD replicas are metastable (in practice, stable), and this study confirms the finding with different systematics.  

In this work we have solved the quark mass gap equation of a global-color model of Coulomb-gauge Hamiltonian of QCD~\cite{Llanes-Estrada:2001bgw}
in which the longitudinal color-density $\rho^a(\vec{x})=\Psi^\dagger_x T^a \Psi_x$  interaction is taken to be a Cornell linear + Coulomb potential, and a current $\Vec{J}^a(\vec{x})=\Psi^\dagger_x T^a\Vec{\alpha} \Psi_x$-current interaction models the gapped transverse gluon exchange:
\begin{align}
    H =\int\! \dif^3 x  \Psi_ {\Vec{x}}^\dagger(-i\vec{\alpha}\cdot \vec{\nabla}+\beta m_q)\Psi_{\Vec{x}}+\dfrac{1}{2}\int\!  \dif^3 x \dif^3 y\left(  -\rho^a_{\vec{x}}V(|\vec{x}-\vec{y}|) \rho^a_{\vec{y}}+ \Vec{J}^a_i(\vec{x})U_{ij}(\vec{x},\vec{y}) \Vec{J}^a_j(\vec{y})\right)
\end{align}
whose longitudinal part $V$ in momentum space, from a pure Yang-Mills variational computation~\cite{Szczepaniak:2001rg}, 
\begin{equation}
    V(p)=
     -\dfrac{8.07}{p^2}\dfrac{\log^{-0.62}\left(\frac{p^2}{m_g^2}+0.82\right)}{\log^{0.8}\left(\frac{p^2}{m_g^2}+1.41\right)} \hspace{0.2cm} \text{if} \hspace{0.1cm} p>m_g;\ \ \ \ \ 
     V(p)=-\dfrac{12.25\, m^{1.93}_g}{p^{3.93}} \hspace{0.2cm} \text{if} \hspace{0.1cm} p<m_g
\end{equation}
 is (numerically) close to   the traditional  $V_{\text{Cornell}}(k)=-4\pi \dfrac{\alpha_s}{k^2}-8\pi\dfrac{\sigma}{k^4} $.
 The transverse part is in turn
\begin{equation}
\label{eq:hyper}
    U_{ij}(\vec{x},\vec{y})=\left(\delta_{ij}-\dfrac{\nabla_i \nabla_j}{\nabla^2}\right)_{\Vec{x}}U(|\vec{x}-\vec{y}|)\ ,
\end{equation}
\begin{equation}
    U(p)= V(p)\hspace{0.2cm}  \text{if} \hspace{0.2cm} p>m_g;\ \ \ \ \ 
     U(p)=-\dfrac{C_h}{p^2+m^2_g} \hspace{0.2cm} \text{if} \hspace{0.2cm} p<m_g
\end{equation}
with  $C_h$ a fixed constant to continuously matching both;
the scale of the theory $m_g=0.6$ GeV.

The trivial vacuum, defined by $b_\lambda|0\rangle=d_\lambda|0\rangle=0$, is related to the BCS constituent quasiparticle vacuum $B_\lambda|\Omega\rangle=D_\lambda|\Omega\rangle=0$ by the transformation
\begin{equation}
\label{eq:vacuum}
\begin{split}
    |\Omega\rangle=\exp \Bigg(-\sum_{\lambda_1,\lambda_2} \int \dfrac{\dif^3 k}{(2\pi)^3} (\vec{\sigma}\cdot \hat{k})_{\lambda_1 \lambda_2} \tan{\dfrac{\theta_k}{2}}
    b_{\lambda_1}^\dagger(\Vec{k})
    d_{\lambda_2}^\dagger(-\Vec{k}) \Bigg)|0\rangle.
    \end{split}
\end{equation}

\begin{figure}
\includegraphics[height=5cm]{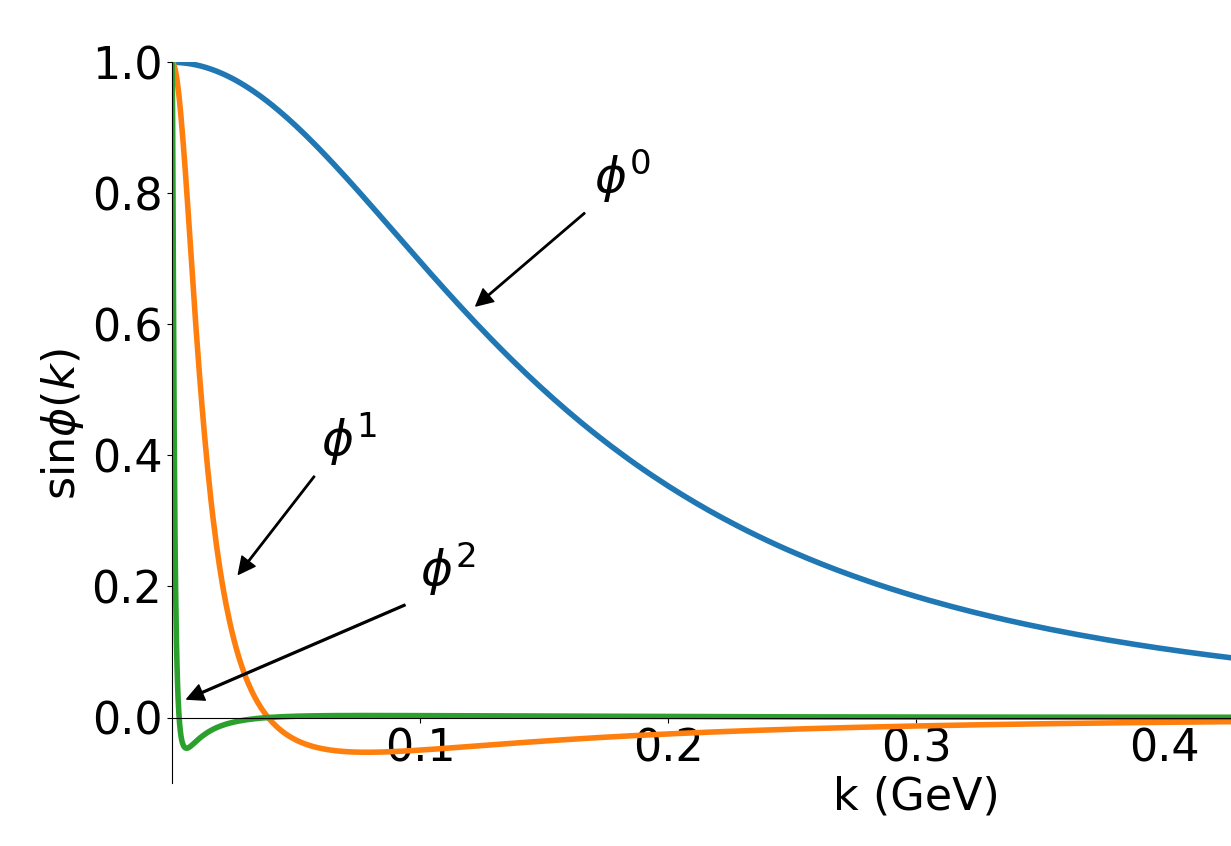}
\includegraphics[height=5cm]{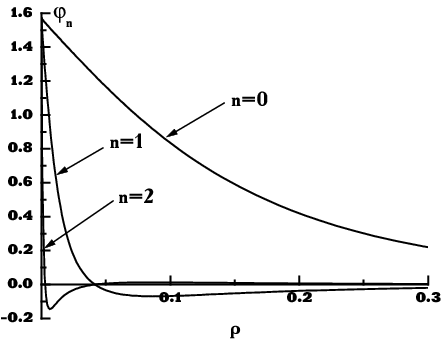}
\caption{\label{solutions}The linear+Coulomb (left) and harmonic oscillator (right~\cite{Bicudo:2019ryc}) solutions to the mass gap equations are very similar and we can confirm the replica picture in which the ground state BCS state $\phi^0$ is accompanied by two excited solutions of the equation. Their number and intensity seems to be tied to the strength of the $\rho_{\rm color}^a \rho_{\rm color}^a$ Coulomb-gauge color-density interaction, independently of the precise kernel details.  }
\end{figure}

The mass-gap equation of this Hamiltonian
$\dfrac{\delta \langle \Omega| H |\Omega\rangle}{\delta \theta_k}=0$
is numerically solved, and its ground state as well as two excited solutions are shown in figure~\ref{solutions}, where they are compared to those of a simple harmonic oscillator. 
It is clear that the basic physics is captured by basically any potential with the correct scale for chiral symmetry breaking yielding a reasonable meson spectrum (the Bogoliubov $\theta$ and BCS $\phi\equiv\varphi$ angles are equal in the chiral limit).

The quark condensates $\langle \bar{\Psi}_x\Psi_x \rangle$ in the BCS ground state $\ar \Omega \ra$, and the two computed replicae $\ar \Omega' \ra$ and $\ar \Omega''\ra$ are respectively $-(178 {\rm MeV})^3$, $(73 {\rm MeV})^3$ and $-(61 {\rm MeV})^3$, indicating a pattern of decreasing chiral symmetry breaking towards the perturbative vacuum (where the condensate vanishes, the pion is no Goldstone boson and it has finite mass).

\begin{figure}
    \hspace{-1cm}
    \includegraphics[width=1.1\columnwidth]{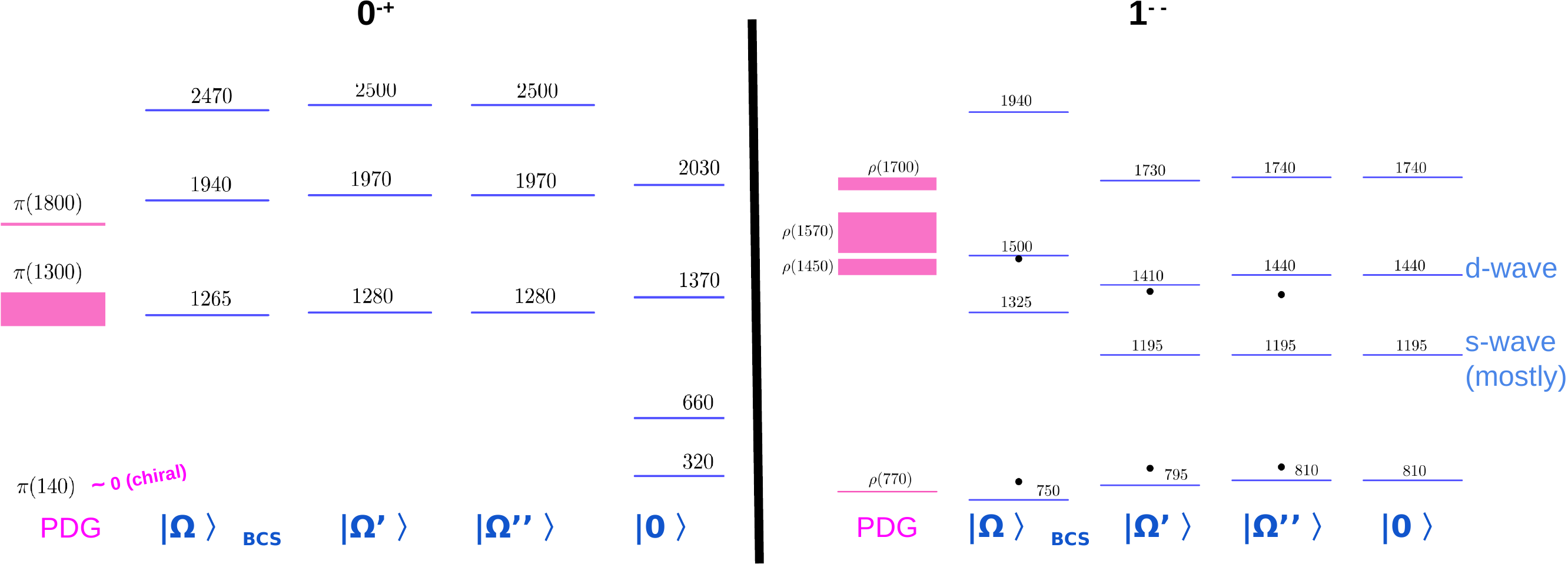} \centering
    \caption{Spectra of pseudoscalar and vector mesons built over the ground-state BCS vacuum, the excited replica vacua and the perturbative (chirally-symmetric, $\theta(k)=0$) state, respectively.  Curiously, the various replicas present similar spectra with $\chi^2$ respect to experiment that only slowly deteriorates. The perturbative vacuum can be seen as a limit of the replica vacua in which the gap angle is exactly zero in the chiral limit, with chiral symmetry not broken
    as shown by the finite pion mass.
    \label{fig:spectrum}
}
\end{figure}

However,  and as shown in figure~\ref{fig:spectrum}, the spectra of pseudoscalar and vector mesons built as $B^\dagger D^\dagger \ar \Omega \ra_i$ in the RPA/instantaneous Bethe-Salpeter approximation, is rather similar for the replicas and the ground state vacuum (and also for the perturbative state $\ar 0\ra$ except for the pion mass that becomes finite), reflecting the rather robust constituent quark model structure.
Of importance for our discussion is that no meson (the scalar ones were also considered for the harmonic oscillator potential~\cite{Bicudo:2019ryc}) has negative mass, that would force the vacuum to be unstable against the emission of meson pairs at the RPA level. We therefore clearly confirm earlier findings and stability.

At the BCS level, however, we find that the Hessian of the vacuum energy minimisation  $F(k,q)=\dfrac{\delta^2 \langle \Omega| H |\Omega\rangle}{\delta \phi_q \delta\phi_k}$ is positive definite only for the BCS ground state $\ar \Omega \ra$. The first replica  $\ar \Omega' \ra$ has one negative eigenvalue indicating a direction of descent towards the ground state, and making it a saddle point; the second replica has two negative eigenvalues and is thus also a saddle point; and the perturbative vacuum $\ar 0 \ra$ has exactly three negative ones, suggesting that for this physically sensible potential strength there are no further replicas to be found (as the system is very nonlinear, we have no Sturm-Liouville like  theorem at hand).

Therefore, we find that, if the function $\langle H\rangle (\theta_k)$ was taken as a classical potential surface $V(\theta_k)$ to be used, for example, as an inflationary potential (but with the scale increased from 1 to $10^{15}$ GeV, of course), the replica vacua would be unstable to specific variations of $\theta_k$.

The interesting twist, however, is that in the quantum theory that variation needs to proceed by the creation of Cooper pairs.
Once the corresponding BCS vacuum has been normalized $\langle \Omega\ar \Omega \ra =
\langle 0 \ar 0\rangle$, transitions between the BCS and perturbative vacuum, or between any of them and the replicas, are suppressed with the exponential of the volume. 
For example, near $\theta_k\simeq 1$,
\begin{equation}
		\braket{\Omega|0}
		= e^{-2V} \left[\alpha_2 \beta \tan{\theta_{\boldsymbol{k}}}|_{\boldsymbol{k}=0} + \bra{0} \int d^3{\boldsymbol{k}} \tan{\theta_{\boldsymbol{k}}} \left\{
		\alpha_0 (bb^\dagger dd^\dagger) 
%		+ \alpha_1(b^\dagger d^\dagger)  
        \right\}\ket{0} +\dots \right]
\end{equation}
in terms of certain coefficient functions $\alpha$, $\beta$ of no concern now.

In an infinite volume we therefore conclude that the replicas are stable also at the BCS level due to the orthogonality between Fock spaces. In fact, we are 
confronted with copies of the entire field theory (inequivalent representations of the Fock space) that lead a totally parallel existence: no transition can happen between two of them in any macroscopic volume. What further physical significance this may have will provide very interesting future investigations.

%%%%%%%%%%%%%%%%%%%%%%%%%%%%%%%%%%%%%%%%%%%%%%%%%%%%%%%%%%%%%%%%%%%%%%%%%%%%%%%%%%%%%%%%%%%%
\acknowledgments
%%%%%%%%%%%%%%%%%%%%%%%%%%%%%%%%%%%%%%%%%%%%%%%%%%%%%%%%%%%%%%%%%%%%%%%%%%%%%%%%%%%%%%%%%%%%
\noindent
This project received funding from the EU’s Horizon 2020 research and innovation programme under grant 824093; spanish MICINN grants PID2019-108655GB-I00, -106080GB-C21; U. Complutense de Madrid research group 910309 \& IPARCOS; portuguese CeFEMA under contract for R\&D Units, strategic project UID/CTM/04540/2019, 
and  FCT project CERN/FIS-COM/0029/2017.

%%%%%%%%%%%%%%%%%%%%%%%%%%%%%%%%%%%%%%%%%%%%%%%%%%%%%%%%%%%%%%%%%%%%%%%%%%%%%%%%%%%%%%%%%%%%

%%%%%%%%%%%%%%%%%%%%%%%%%%%%%%%%%%%%%%%%%%%%%%%%%%%%%%%%%%%%%%%%%%%%%%%%%%%%%%%%%%%%%%%%%%%%

\end{document}